\documentclass[aps,twocolumn,superscriptaddress,groupedaddress,10pt,letterpaper]{revtex4} 
\usepackage{graphicx} 
\usepackage{dcolumn} 
\usepackage{bm} 
\usepackage{amssymb} 
\usepackage{amsmath}
\usepackage{subcaption}
\usepackage{booktabs}
\usepackage{bbold}
\usepackage{blindtext}
\usepackage[utf8]{inputenc}
\usepackage{float}
\usepackage{color}
\usepackage{mwe}
\usepackage{multirow}
\usepackage{braket}
\usepackage{caption}
\usepackage{algorithm}
\usepackage{algpseudocode}
\usepackage{diagbox}
\usepackage[dvipsnames]{xcolor}
\newcommand{\CUAaddress}{Harvard-MIT Center for Ultracold Atoms, Cambridge, Massachusetts 02138, USA}
\newcommand{\HarvardPhysicsAddress}{Department of Physics, Harvard University, Cambridge, Massachusetts 02138, USA}
\newcommand{\HarvardChemistryaddress}{Department of Chemistry and Chemical Biology, Harvard University, Cambridge, Massachusetts 02138, USA}
\newcommand{\QuEraaddress}{QuEra Computing, 1284 Soldiers Field Road, Boston, MA 02135, USA}
\newcommand{\Arforaddress}{Institut für Experimentalphysik, Universität Innsbruck, Technikerstr. 25, 6020 Innsbruck, Austria}

\begin{document}

\title{Quantum interference and entanglement in ultracold atom-exchange reactions}

\author{Yi-Xiang~Liu}
\thanks{These authors contributed equally to this work.}
\affiliation{\HarvardChemistryaddress} 
\affiliation{\CUAaddress}

\author{Lingbang~Zhu}
\thanks{These authors contributed equally to this work.}
\affiliation{\HarvardChemistryaddress}
\affiliation{\CUAaddress}

\author{Jeshurun~Luke}
\affiliation{\HarvardChemistryaddress}
\affiliation{\CUAaddress}

\author{J.~J.~Arfor~Houwman}
\affiliation{\Arforaddress}
\affiliation{\HarvardChemistryaddress}

\author{Mark C.~Babin}
\affiliation{\HarvardChemistryaddress}
\affiliation{\CUAaddress}

\author{Ming-Guang~Hu}
\thanks{Current address: \QuEraaddress}
\affiliation{\HarvardChemistryaddress}
\affiliation{\CUAaddress}

\author{Kang-Kuen~Ni}
\email{ni@chemistry.harvard.edu}
\affiliation{\HarvardChemistryaddress}
\affiliation{\HarvardPhysicsAddress}
\affiliation{\CUAaddress}
\date{\today}

\begin{abstract} 
Coherent superpositions and entanglement are hallmarks of quantum mechanics, but they are fragile and can easily be perturbed by their environment. Selected isolated physical systems can maintain coherence and generate entanglement using well-controlled interactions. Chemical reactions, where bonds break and form, are highly dynamic quantum processes. A fundamental question is whether coherence can be preserved in chemical reactions and then harnessed to generate entangled products. Here we investigate this question by studying the 2KRb $\rightarrow$ K$_2$ + Rb$_2$ reaction at 500 nK, focusing on the the nuclear spin degrees of freedom. We prepare the initial nuclear spins in KRb in an entangled state and characterize the preserved coherence in nuclear spin wavefunction after the reaction. The data are consistent with full coherence at the end of the reaction. This suggests that entanglement can be prepared within the reactants, followed by a chemical reaction that produces separate, entangled molecules. We additionally demonstrate control of the reaction product state distribution by deliberately decohering the reactants.
\end{abstract}

\maketitle

\begin{figure*}
\includegraphics[width= \textwidth]{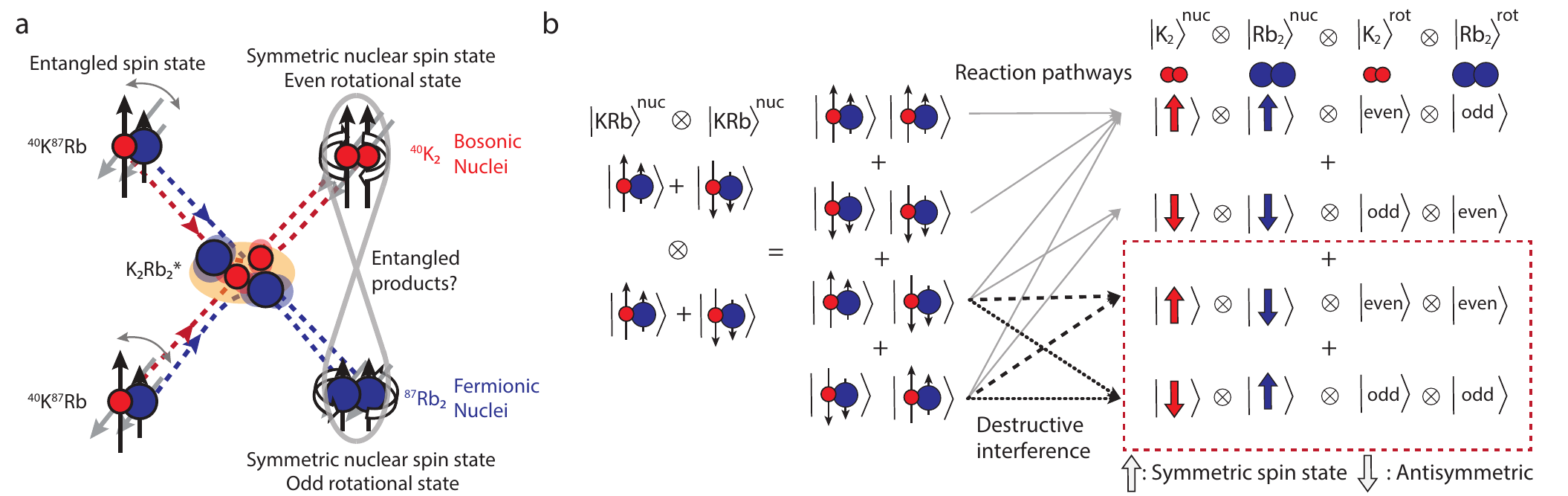}
\caption{\textbf{Atom-exchange reaction with reactant molecules prepared in entangled nuclear spin states yield distinct product outcomes depending on whether the reaction preserves coherence or not.} \textbf{a,} Reaction of interest. Would entangled nuclear spins in the reactant KRb molecule result in entangled products of K$_2$ and Rb$_2$? Because of the exchange symmetry of indistinguishable nuclei in homonuclear diatomic molecules (K$_2$ and Rb$_2$), the symmetry of the nuclear spins can be probed via the parity of the rotation states. \textbf{b,} If the nuclear spin wavefunction stays coherent throughout the reaction, reaction pathways destructively interfere for the $\ket{S,A,e,e}$ and $\ket{A,S,o,o}$ channels. If the nuclear spin wavefunction decoheres into a statistical mixture, either before or during the chemical reaction, no interference between the bottom two paths exists and all four product channels will be populated.}
\label{fig:scheme}
\end{figure*}

Observing and exploiting coherence in reactions has been a long-standing goal in chemistry. Coherence, often manifested through interference, has been observed in reactions of light species such as HD, HF, and D$_2$ in molecular beam experiments, where comparison with full quantum scattering calculations is needed to identify the signatures of interference~\cite{jambrina2015quantum,xie2020quantum,chen2021quantum}. 

Coherence can also be leveraged to control the product state distribution using interferences between reaction pathways~\cite{shapiro1996coherent,kale2021constructive}. Early experimental success has been shown in light-induced processes, such as photo-association~\cite{blasing2018observation}, photo-ionization~\cite{zhu1995coherent}, and photo-dissociation~\cite{sheehy1995phase,shnitman1996experimental}. However, experimental realization of controlled reactive scattering using coherent interference remains elusive. A key underlying question is whether quantum coherence survives reactive bimolecular collisions, given the complex quantum dynamics. If coherence can be preserved, these chemical reactions may not only be controlled, but also harnessed as a resource for quantum science. 

To answer this question, we turn to ultracold molecular systems where exquisite control of the quantum state of molecules, developed over the last two decades, allows for preparation in a single quantum state at $\mathrm{\mu}$K temperature or lower~\cite{ni2008high,danzl2008quantum,lang2008ultracold,cheuk2018lambda,tarbutt2019laser,wu2021high,langin2021polarization}. Specifically, we choose $^{40}$K$^{87}$Rb bialkali molecules as our platform, as the atom-exchange chemical reaction,  2KRb $\rightarrow$ K$_2$ + Rb$_2$, has been observed ~\cite{hu2019direct} and studied extensively~\cite{ospelkaus2010quantum,liu2020photo}.

To probe correlation of the reaction outcomes, it is crucial to identify the quantum states of products from individual events. For this purpose, we take advantage of a previously developed technique of coincidence detection, where pairs of products are simultaneously and state-selectively ionized, probed using velocity map imaging, and filtered based on momentum conservation~\cite{liu2021precision}. 

Such coincident detection was used to characterize the 2KRb $\rightarrow$ K$_2$ + Rb$_2$ process with full state-to-state resolution. Due to the modest exothermicity of the KRb reaction ($\approx$~10~cm$^{-1}$), only rotational excitations of the product molecules are energetically allowed, constraining the product state space for a comprehensive study. The results revealed a mostly statistical distribution of the product rotational states, suggesting underlying chaotic dynamics and implying the difficulty of preserving coherence in the rotational degree of freedom~\cite{liu2021precision}. Nevertheless, reaction products have been controlled by reactant nuclear spins~\cite{hu2021nuclear}, suggesting that nuclear spins play a passive role in such a reaction. This positions the nuclear spin degree of freedom as a promising candidate to study quantum coherence and entanglement within chemical reactions. We note that the role of coherence was not explored in Ref.~\cite{hu2021nuclear} as the state-resolved reaction outcomes were averaged rather than coincidentally detected. 

In this study, we investigate whether coherence is maintained within the nuclear spin degree of freedom throughout an atom-exchange reaction. As illustrated in Fig.~\ref{fig:scheme}, we design an experiment with reactants prepared in an entangled nuclear spin state and measure the outcome to distinguish between coherent and incoherent processes. If coherence is maintained, it will become evident in the final product nuclear spin state distribution. However, detecting the product molecule nuclear spin state spectroscopically is generally challenging due to small energy splittings compared to the natural linewidth of the transitions and Doppler shifts from the fast-flying products. Instead, we indirectly probe the nuclear spin degree of freedom making use of the fact that the symmetry of the nuclear spin states and the parity of the orbital wavefunction respect the exchange symmetry of the two indistinguishable nuclei~\cite{sakurai1995modern} in the homonuclear diatomic product molecules K$_2$ and Rb$_2$. Specifically, the nucleus of $^{40}$K ($^{87}$Rb) is Bosonic (Fermionic) and thus the total nuclear wavefunction of K$_2$ (Rb$_2$), which can be factored into a spin part and a spatial part, is symmetric (antisymmetric) under exchange~\cite{hu2021nuclear} (Fig.~\ref{fig:scheme}a). By probing the parity, even or odd, of the rotational states of the outgoing products, we obtain the symmetry of the product nuclear spins. 

Previous work suggests that the nuclear spin dynamics in the reaction process can be described by transforming the two nuclear spins in K or Rb (in two separate KRb molecules) from the uncoupled basis to the coupled basis in K$_2$ or Rb$_2$~\cite{hu2021nuclear}. Assuming that the nuclear spin wavefunctions remain coherent throughout the reaction, coupling two K or two Rb with the same spin projection leads only to a symmetric total spin state $|S\rangle$. Coupling two K or two Rb with different spin projections results in a superposition of $|S\rangle$ and $|A\rangle$ ($A$ is antisymmetric), where the relative phase depends on the coupling order, causing interference between different reaction pathways [Appendix \ref{appendix:pure_and_mixed}]. Consequently, as illustrated in Fig.~\ref{fig:scheme}b, destructive interference leads to complete suppression of the $\ket{\mathrm{K}_2}^\mathrm{nuc}\otimes \ket{\mathrm{Rb}_2}^\mathrm{nuc}$ = $\ket{S}\otimes\ket{A}$ and the $\ket{A}\otimes\ket{S}$ channels. Due to the spin-rotation correspondence of the K$_2$ and Rb$_2$ nuclei, the product states $|e,e\rangle$ and $|o,o\rangle$, where $e$ is an even-numbered and $o$ is an odd-numbered rotation state, are also completely suppressed, and the population is entirely in the $|e,o\rangle$ and $|o,e\rangle$ channels. Both the spin and the rotational components of the K$_2$ and Rb$_2$ molecule wavefunctions become entangled, as they inherit the entanglement from the reactants \footnote{If the initial KRb nuclear spin state is a product state of K and Rb nuclear spins and remains pure, regardless of their specific components, the resulting K$_2$ and Rb$_2$ molecules will always exhibit $\ket{e,o}$ due to symmetry protection.}. In contrast, if the nuclear spin states are no longer pure, due to decoherence occurring either before or during the reaction, the two K and Rb nuclei are no longer indistinguishable and the symmetry protection is lost. As a consequence, $\ket{S,A}(|e,e\rangle)$ and $\ket{A,S}(|o,o\rangle)$ channels will be populated in addition to the $\ket{S,S}(|e,o\rangle)$ and $\ket{A,A}(|o,e\rangle)$ channels. This provides a feasible way to test whether coherence is preserved. 

\begin{figure*}
\centering
\includegraphics[width=0.95\textwidth]{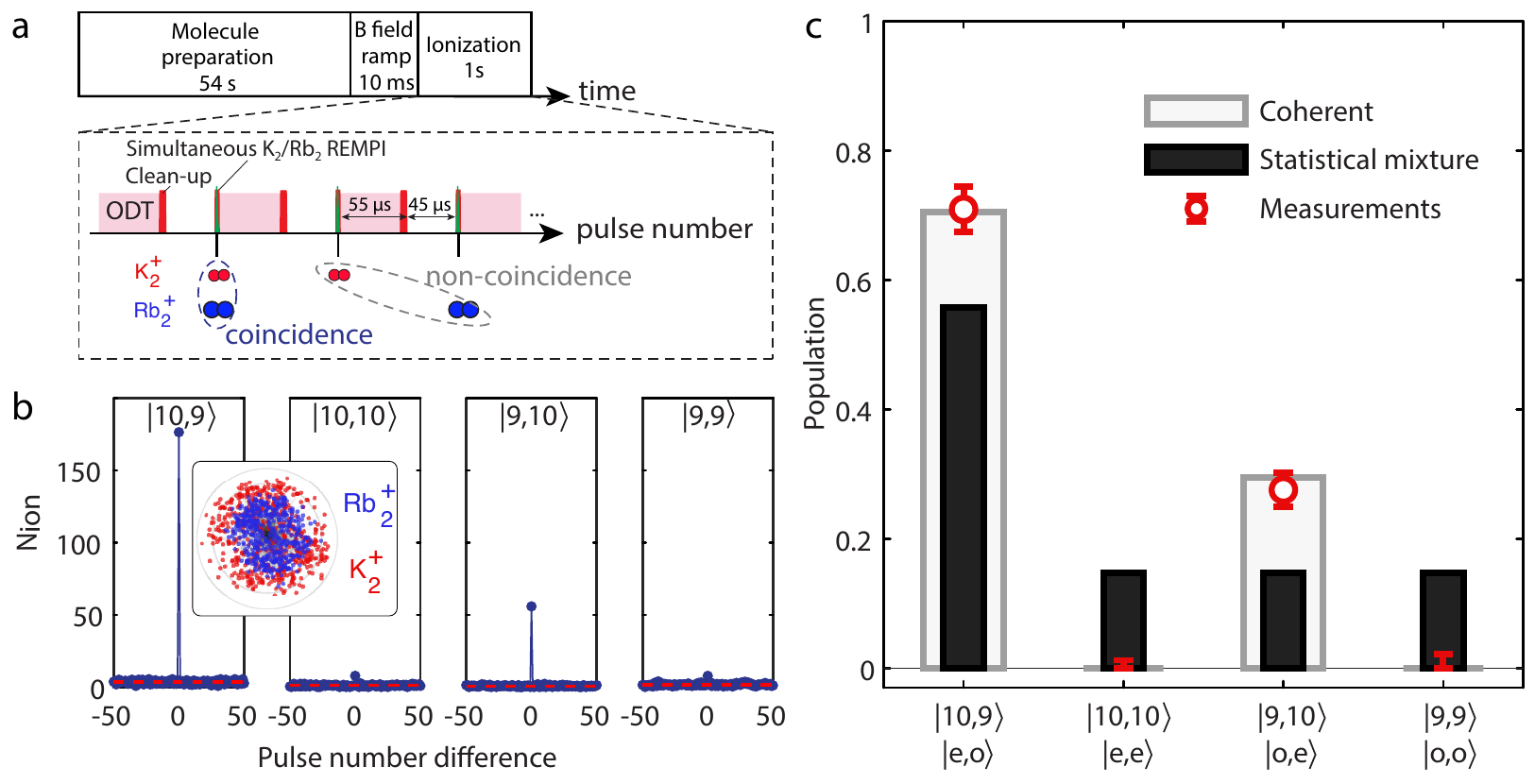}
\caption{\textbf{Coincidence detection of coherent reaction products.} \textbf{a,} Experimental sequence highlighting product ionization detection timing. \textbf{b,} K$_2^+$-Rb$_2^+$ pair counts per 1000 cycles after momentum filtering along all three directions as a function of pulse number difference between the two ions. Pulse number difference of 0 means that the K$_2^+$ and Rb$_2^+$ are generated by simultaneous ionization. (Inset) $|10,9\rangle$ pair velocity map image on the XY plane. \textbf{c,} The normalized measured population of the products (red data) overlaid on the expected K$_2$ and Rb$_2$ rotational state population distributions: if the nuclear spin wavefunction remains pure, the population distribution follows the grey bars; if the nuclear spin wavefunction fully dephases, the expected distribution are represented by the black solid bars. The $|10,9\rangle$ and $|9,10\rangle$ error bars are propagated from the shot noise; the $|9,9\rangle$ and $|10,10\rangle$ ranges are propagated from the 68\% confidence intervals of the count rates after background subtraction using Bayesian analysis. The measured results align well with the fully coherent case, indicating that the nuclear spins stay coherent throughout the reaction.}
\label{fig:coherent}
\end{figure*}

Our experiment starts with the preparation of ultracold rovibrational ground-state $^{40}$K$^{87}$Rb molecules in a single hyperfine state at 500~nK. The molecular gas is trapped in an optical dipole trap (ODT) in an apparatus with ion spectrometry and velocity map imaging~\cite{liu2020probing}. The entangled KRb state can be prepared relatively easily by magnetic field ramping due to the interaction between the two nuclei in KRb [Appendix~\ref{appendix:Hamiltonian}]. We start with KRb in a product state $\ket{\psi_{i}} = \ket{N,m_N,m_{I}^{\mathrm{K}},m_{I}^\mathrm{Rb}} = \ket{0,0,-4,1/2}$ at 542 G, where $N$ is the rotational quantum number, $m_N$ is the rotation projection, and $m_{I}^{\mathrm{K}}$ ($m_{I}^{\mathrm{Rb}}$) is the K (Rb) nuclear spin projection along the quantization axis. We then ramp the magnetic field to 5 G within 10 ms. Throughout the field ramp, the KRb molecules remain in an energy eigenstate [Appendix~\ref{appendix:Hamiltonian}] and end in the state 
\begin{align}
\ket{\psi_{f}} = ~&\alpha \ket{0,0,-4,1/2} + \beta \ket{0,0,-3,-1/2} \nonumber \\
&+ \gamma \ket{0,0,-2,-3/2}
\end{align}
where $(\alpha,\beta,\gamma) = (0.595 , 0.719, 0.359)$ at 5~G and the K nuclear spin and the Rb nuclear spin are entangled. The 2KRb $\rightarrow$ K$_2$ + Rb$_2$ reaction occurs continuously through collisions. We strobe the ODT on and off at a 10-kHz repetition rate which allows us to probe products formed in the dark, thereby minimizing any perturbations from the ODT. The reaction products formed when the ODT is on are removed by clean-up pulses (Fig.~\ref{fig:coherent}a).

If the coherence is preserved, as shown in Fig.~\ref{fig:scheme}b, we expect the populations in the different rotational parity channels to be
\begin{align}
P_C \equiv &\begin{bmatrix}
P_{eo} & P_{ee} \\
P_{oo} & P_{oe} 
\end{bmatrix}
= \begin{bmatrix}
1 - x & 0 \\
0 & x \\
\end{bmatrix}
= \begin{bmatrix}
0.7049 & 0 \\
0 & 0.2951\\
\end{bmatrix},
\label{eqn:coherent}
\end{align}
where $x = |\alpha\beta|^{2} + |\beta\gamma|^{2} + |\alpha\gamma|^{2}$ [Appendix~\ref{appendix:pure_and_mixed}]. Here $P_{eo}$ refers to the population fraction with even K$_2$ and odd Rb$_2$ rotational quantum numbers. $P_{oe}$, $P_{ee}$, and $P_{oo}$ are defined analogously, and all four populations add to one.
If the nuclear spin state decoheres into a statistical mixture, we expect

\begin{align}
P_S\equiv\begin{bmatrix}
1 - \frac{3}{2}x & \frac{1}{2}x \\
\frac{1}{2}x & \frac{1}{2}x
\end{bmatrix}
= \begin{bmatrix}
0.5575 & 0.1475 \\
0.1475 & 0.1475\\
\end{bmatrix}.
\label{eqn:incoherent}
\end{align}

Coincident detection is required to distinguish between the coherent and statistical mixture outcomes, as non-coincident detection would only be sensitive to the sum of the other product (e.g. $P_{N_{\mathrm{K}_2}=even} =P_{eo}+P_{ee} $). Experimentally, we restrict ourselves to the state pairs $|N_{\mathrm{K}_2},N_\mathrm{Rb_2}\rangle$ = $|10,9\rangle$, $\ket{10,10}$, $\ket{9,10}$ and $\ket{9,9}$ due to their higher signal compared to other channels. Additionally, the degeneracy of the $|10,9\rangle$ pair is well described by the statistical model~\cite{liu2021precision}, which is essential for proper normalization. 

We probe the population of molecules in each rotation state using state-selective resonance-enhanced multiphoton ionization (REMPI) of K$_{2}$ and Rb$_{2}$ following Ref.~\cite{hu2021nuclear,liu2021precision} [Appendix~\ref{appendix:rempi}] and a timing sequence shown in Fig.~\ref{fig:coherent}a. For each experimental cycle (cycle time of 55 seconds), we probe the reaction products for 1 s. We accumulate data for $\sim$ 3000 cycles ($\sim 46$ hours) for each state pair. Momentum conservation ensures that the total momentum of the K$_2$ and Rb$_2$ molecules originating from the same reaction events remains zero, as the reactants start with near zero momentum. We filter the momentum along the X and Y directions using position information from velocity map imaging and additionally along Z using the time-of-flight for all collected reaction events to extract coincidence counts of each product state pairs [Appendix~\ref{appendix:coi}]. The accumulated coincidence count rates of the four state pairs (Fig.~\ref{fig:coherent}b) are 
$\begin{bmatrix}
N_{|10,9\rangle} & N_{|10,10\rangle} \\
N_{|9,9\rangle} & N_{|9,10\rangle} 
\end{bmatrix}$ = $\begin{bmatrix}
173.1 \pm 8 & 6.7 \pm 1.8 \\
6.1 \pm 1.8 & 55.2 \pm 4.2 
\end{bmatrix}$,

where error bars represent shot noise. 
The coincidence counts of $|10,10\rangle$ and $|9,9\rangle$ could represent decoherence or emerge from other source, such as the ionization of other state pairs. This may occur when one product is ionized through state-selective REMPI and the other product is off-resonantly ionized by two 532~nm photons, which could register as a coincident count during data accumulation. 

To quantify such a background contribution, we detect the $|10,10\rangle$ channel at 50 G, where the KRb is dominantly in $\ket{0,0,-4,1/2}$ (with 0.994 population) and we do not expect any reaction outcome in $|10,10\rangle$. In this background measurement, we found a coincidence count rate of 7.1 $\pm$ 1.7, which can account for the entire contribution to the measured $|10,10\rangle$ and $|9,9\rangle$ counts at 5~G. Additionally, we look for counts that come from both products being off-resonantly ionized by turning off the resonant legs of the REMPI beams and find them negligible.

For the $|10,9\rangle$ and $|9,10\rangle$ channels, we simply subtract this background source from the measured signal. For the $|10,10\rangle$ and $|9,9\rangle$ channels, because the measured signal is lower than the measured background due to random fluctuations, we use Bayesian analysis \cite{little1982statistical} to find the 68\% confidence intervals. We then normalize all channels, accounting for the velocity factors and channel degeneracies [Appendix~\ref{appendix:normalization}], to extract the branching ratios
$\begin{bmatrix}
0.709 \pm 0.035 & [0, 0.007] \\
[0,0.012] & 0.276 \pm 0.026 
\end{bmatrix}$.
These measurements, shown in Fig.~\ref{fig:coherent}c, are near the coherent values of $P_C$ (Eq.~\ref{eqn:coherent}).

To further quantify the degree of coherence that is preserved in this reaction, we assume a dephasing model (as previous work in Ref.~\cite{hu2021nuclear} established population conservation) where all the off-diagonal terms in the density matrix decay by an equal amount. As such we multiply the off-diagonals of the initial pure state KRb + KRb spin density matrix by a common factor $\Gamma$ to find the decohered spin density matrix of the reaction products
\begin{align}
\rho_\mathrm{\mathrm{K_2},\mathrm{Rb_2}} &= \Gamma U \rho_\mathrm{KRb}\otimes\rho_\mathrm{KRb} U^{\dagger} \nonumber \\
&+ (1-\Gamma)U \left( \rho_\mathrm{KRb}\otimes\rho_\mathrm{KRb}\right)_\mathrm{diag} U^{\dagger} ,
\end{align} 
where $\rho_\mathrm{KRb}=\ket{\psi_f}\bra{\psi_f}$, $U$ is the unitary transformation that transforms the two K and Rb nuclear spins from uncoupled basis to coupled basis and $\left( \rho_\mathrm{KRb}\otimes\rho_\mathrm{KRb}\right)_\mathrm{diag}$ is the diagonal part of the pure state density matrix $\rho_\mathrm{KRb}\otimes\rho_\mathrm{KRb}$. With these assumptions, the final branching ratio can be described by $\Gamma P_C+(1-\Gamma)P_S$. From Bayesian analysis [Appendix~\ref{appendix:bayesian}], we find the result, in the space of detected products, is consistent with full coherence ($\Gamma=1$) and no lower than 0.9014 (95\% confidence), limited by data statistics. These results show that the nuclear spin wavefunction remains coherent throughout the reaction, from which we can further infer that the nuclear spin states of K$_2$ and Rb$_2$ molecules, along with their rotational states, are entangled. An intriguing question regarding the lifetime of the entanglement after the products separate is beyond the scope of this study.

\begin{figure*}
\includegraphics[width=0.95\textwidth]{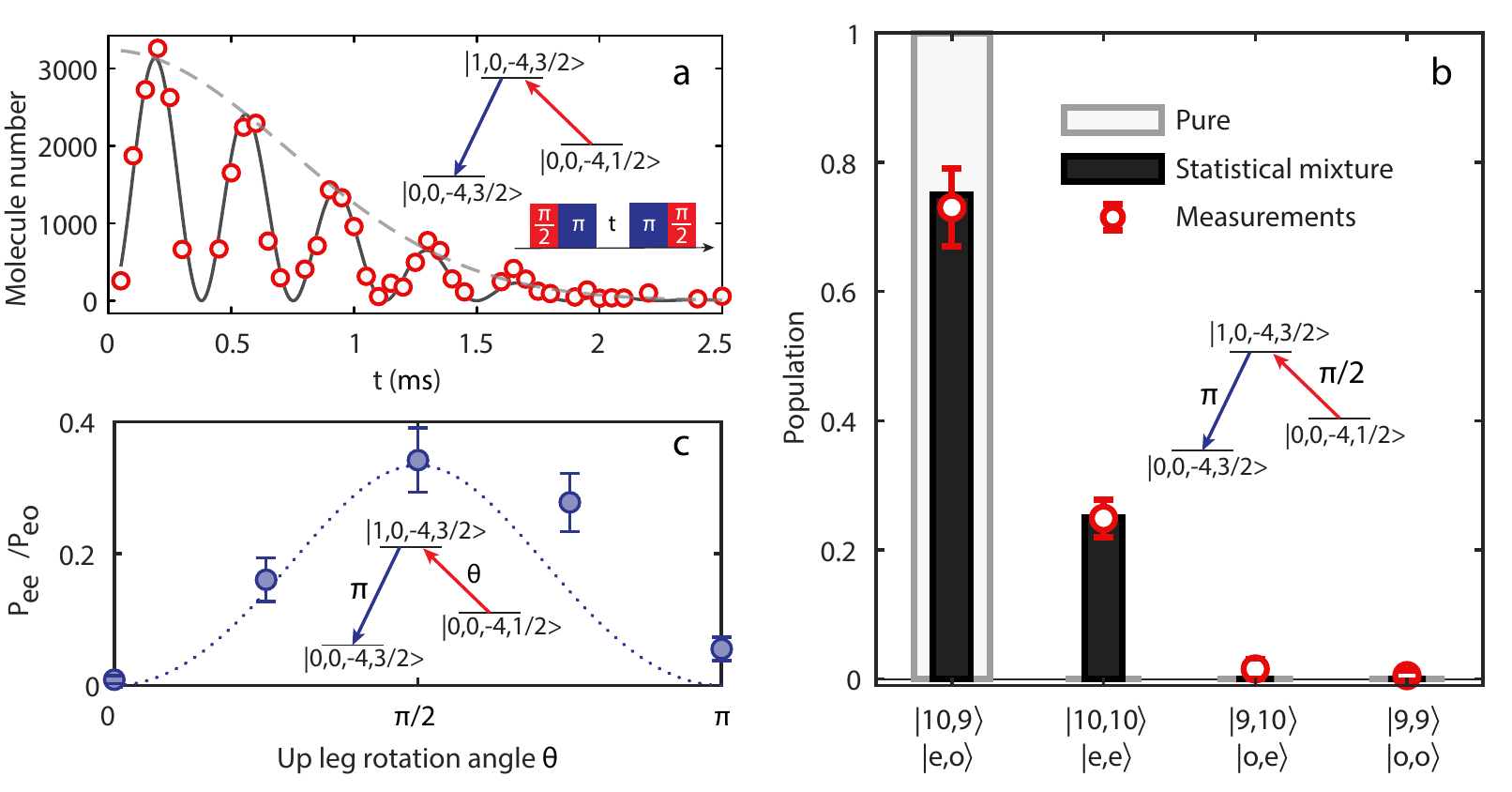}
\caption{\textbf{ Reaction with reactant molecules prepared in a statistical mixture.} \textbf{a,} Ramsey spectroscopy of KRb molecules in the superposition of two hyperfine states, $|0,0,-4,1/2\rangle$ and $|0,0,-4,3/2\rangle$, prepared by two consecutive microwave pulses (scheme and timing shown in the inset). 
The solid line is a fitting of molecule number to a damped oscillation with a Gaussian decay envelope. The dashed line is the fitted envelope $\exp(-(t/T_2)^2)$, from which $T_2$ = 1.03(6)~ms is extracted. 
\textbf{b,} Comparison of the measured population of the four state pairs (open red circle) vs the expectation for the pure and statistical mixture outcomes. The data is taken with reactants prepared in a 50-50 mixture of $|0,0,-4,1/2\rangle$ and $|0,0,-4,3/2\rangle$. Error bars are propagated from the shot noise of each state pair. The black bars are the expected population for a completely decohered state, while the gray bars show the expected population distribution for a pure initial state $\frac{1}{\sqrt{2}}|0,0,-4,1/2\rangle + \frac{1}{\sqrt{2}}|0,0,-4,3/2\rangle$. \textbf{c,} Controlling the population between $\ket{e,e}$ and $\ket{e,o}$ channels by tuning the population ratio in the statistical mixture of $|0,0,-4,1/2\rangle$ and $|0,0,-4,3/2\rangle$. The population of the $|0,0,-4,1/2\rangle$ state is $\cos^2(\frac{\theta}{2})$ and the population of $|0,0,-4,3/2\rangle$ is $\sin^2(\frac{\theta}{2})$, where $\theta$ is the up leg rotation angle. The dotted line is the theory calculation of the $P_{ee}/P_{eo}$ following Eq.~\ref{eqn:theta} overlaid on data (blue circles), where the error bars are propagated from the shot noise on the $|10,10\rangle$ and $|10,9\rangle$ pairs. The deviation of the measurements from the dotted line at large rotation angles could be attributed to imperfect state preparation.}
\label{fig:incoherent}
\end{figure*}

As we find the initial reactant state $\ket{\psi_f}$ to berobust to decoherence, both from experimental noise and chemical reaction, we cannot contrast the coherent outcome to that of a statistical mixture by letting the state decohere. Thus, we deliberately prepare the reactants in a statistical mixture of two energy eigenstates. Specifically, we drive microwave transitions to prepare the KRb molecules in a superposition of two energy eigenstates $\ket{0,0,-4,1/2}$ and $\ket{0,0,-4,3/2}$ via the intermediate state $\ket{1,0,-4,3/2}$ using sequential microwave pulses (Fig.~\ref{fig:incoherent}a inset). The ambient magnetic and electric field are set to 200~G and 424~V/cm, where the microwave transitions are sufficiently isolated from neighboring ones. By applying an up-leg pulse with rotation angle $\theta$ on the Bloch sphere and a down-leg $\pi$ pulse, we create a superposition state $\cos(\theta/2)\ket{0,0,-4,1/2} + \sin(\theta/2)\ket{0,0,-4,3/2}$. We characterize the coherence decay using Ramsey spectroscopy and find a Gaussian decay envelope with a 1.03(6) ms time constant (Fig.~\ref{fig:incoherent}a). This time scale is much shorter than the p-wave collision time of $\sim$ 250 ms [Appendix Fig.~\ref{fig:molecule_lifetime}] for indistinguishable Fermionic KRb molecules measured for our experimental conditions. Therefore, the reaction process is dominated by s-wave reactions between distinguishable KRb molecules, which proceed $\sim$100 times faster than the p-wave collision rate~\cite{ospelkaus2010quantum}. Following the procedure used to produce Eq. \ref{eqn:coherent} and \ref{eqn:incoherent}, it can be shown that the expected population distribution for this statistical mixture is
\begin{align}
\begin{bmatrix}
P_{eo} & P_{ee} \\
P_{oo} & P_{oe} 
\end{bmatrix} = 
\begin{bmatrix}
1-y & y \\
0 & 0
\end{bmatrix},
\label{eqn:theta}
\end{align}
where $y=\sin^2(\frac{\theta}{2})\cos^2 (\frac{\theta}{2})$ and the emergence of $\ket{e,e}$ is a result of decoherence. For a 50-50 statistical mixture ($\theta = \pi/2$) of $\ket{0,0,-4,1/2}$ and $\ket{0,0,-4,3/2}$, the measured branching ratio of the four states is shown in Fig.~\ref{fig:incoherent}b, aligning well with the expected values in Eq.~\ref{eqn:theta} and demonstrating control of the reaction products using decoherence. Furthermore, by tuning the population ratio between the $\ket{0,0,-4,1/2}$ and $\ket{0,0,-4,3/2}$ states, the reaction product distribution agrees with the theoretical predictions (Fig.~\ref{fig:incoherent}c). 

In summary, we investigate the fundamental question of whether coherence can be preserved in reactive scattering using a model atom-exchange reaction that proceeds in a well-isolated gas phase setup. We examine specifically the nuclear spin degree of freedom by preparing entangled nuclear spins in reactants and measuring product outcomes. Similar to Hong-Ou-Mandel interference of indistinguishable particles~\cite{HOM,Kaufman2014}, the coherent phase information is mapped onto the outcome population distribution, manifested as destructive interference of the $|e,e\rangle$ and $|o,o\rangle$ channels. By measuring populations of each parity channel, we observe that coherence of the nuclear spin wavefunction is preserved, which suggests that the products are in an entangled state $\sqrt{P_{eo}}|e,o\rangle+\sqrt{P_{oe}}|o,e\rangle$.  However, direct characterization of the relative phase in the product quantum state would require future measurements in a different basis~\cite{sakurai1995modern}.
 
While this work is reminiscent of hyperfine spin entanglement generated in atomic collisions~\cite{sorensen2001many,esteve2008squeezing,hamley2012spin}, the persistence of nuclear spin coherence in the presence of short-range interactions during chemical reactions, which typically couple different degrees of freedom, is a pleasant surprise. Our work shows that a reaction that exchanges atom partners can be a resource to generate entangled pairs, allowing trivial entanglement to be broken up in parting molecules -- a strategy that may be generalized in other chemical processes. Finally, our findings precede the study of coherence in reactions under wet and warm conditions, which may be of interest for a wide range of chemical phenomena, including in the brain~\cite{fisher2015quantum}. 

\section*{Acknowledgement}
This work is funded by NSF-EAGER through Grant~CHE-2332539, the Gordon and Betty Moore Foundation through  GBMF11558, and the David and Lucile Packard Foundation. We thank T. Rosenband for Bayesian data analysis and the BEC2023 community for discussions. K.-K.N. thanks C. Lieber for enthusiastic support of fundamental research even in difficult times. 

\onecolumngrid
\appendix

\subsection{Expected coincident product pair ratios with pure and fully mixed reactant states} \label{appendix:pure_and_mixed}

Here we discuss the difference of the reaction outcomes in two scenarios, one is when KRb molecules are prepared in a pure superposition quantum state and the other is when KRb reactants start in a fully mixed state.

In the first case, the KRb molecules start in a superposition state $\ket{\psi_f} = \alpha\ket{-4,1/2} + \beta\ket{-3,-1/2} + \gamma\ket{-2,-3/2}$, where the coefficients ($\alpha$,$\beta$,$\gamma$)=(0.595,0.719,0.359) at B = 5 G \cite{hu2021nuclear}. In a KRb + KRb reaction event, the wavefunction of the 4-atom composite could be written as $\ket{\psi_f} \otimes \ket{\psi_f} = \alpha^{2} \ket{-4,1/2} \otimes \ket{-4,1/2} + \beta^{2} \ket{-3,-1/2} \otimes \ket{-3,-1/2} + \gamma^{2} \ket{-2,-3/2} \otimes \ket{-2,-3/2} + \alpha\beta (\ket{-4,1/2} \otimes \ket{-3,-1/2} + \ket{-3,-1/2} \otimes \ket{-4,1/2}) + \alpha\gamma (\ket{-4,1/2} \otimes \ket{-2,-3/2} + \ket{-2,-3/2} \otimes \ket{-4,1/2}) + \beta\gamma (\ket{-3,-1/2} \otimes \ket{-2,-3/2} + \ket{-2,-3/2} \otimes \ket{-3,-1/2})$. Here, we use the $\ket{m_{\mathrm{K}1}, m_{\mathrm{Rb}1}} \otimes \ket{m_{\mathrm{K}2}, m_{\mathrm{Rb}2}}$ basis to represent the quantum state of the four atoms and drop the quantum number $I_\mathrm{K} = 4$ and $I_\mathrm{Rb} = 3/2$. After the reaction, the two K atoms and Rb atoms pair up to form $\mathrm{K_{2}}$ and $\mathrm{Rb_{2}}$. Therefore, we perform a basis transformation into the coupled basis $\ket{I_{\mathrm{K_2}}, m_{\mathrm{K_2}}} \otimes \ket{I_{\mathrm{Rb_2}}, m_{\mathrm{Rb_2}}}$. The symmetry of the product nuclear spin state can be determined by calculating $I_{\mathrm{K(Rb)}} + I_{\mathrm{K(Rb)}} - I_{\mathrm{K_{2}(Rb_{2}})}$, which is symmetric if the quantity is even and antisymmetric if the quantity is odd.

\begingroup
\allowdisplaybreaks
\begin{align}
   &\psi_\mathrm{K_2,Rb_2}\\
    &=\alpha^2|8,-8\rangle\otimes\left(\sqrt{\frac{3}{5}}|3,1\rangle-\sqrt{\frac{2}{5}}|1,1\rangle\right) 
        \\
    &+\beta^2\left(-\sqrt{\frac{7}{15}}|6,-6\rangle+\sqrt{\frac{8}{15}}|8,-6\rangle\right)\otimes
        \\
    &\left(\sqrt{\frac{3}{5}}|3,-1\rangle-\sqrt{\frac{2}{5}}|1,-1\rangle\right)
        \\
    &+\gamma^2\left(\sqrt{\frac{45}{143}}|4,-4\rangle-\sqrt{\frac{14}{55}}|6,-4\rangle+\sqrt{\frac{28}{65}}|8,-4\rangle\right)\otimes
        \\
    &|3,-3\rangle	
        \\
    &+\alpha\beta\left(-\sqrt{\frac{1}{2}}|7,-7\rangle+\sqrt{\frac{1}{2}}|8,-7\rangle\right) \otimes \label{destructive_interference_1}
        \\
    &\left(-\frac{1}{2}|0,0\rangle-\frac{1}{2\sqrt{5}}|1,0\rangle
    +\frac{1}{2}|2,0\rangle+\frac{3}{2\sqrt{5}}|3,0\rangle\right) \label{destructive_interference_2}
        \\ 
    &+\alpha\beta\left(\sqrt{\frac{1}{2}}|7,-7\rangle+\sqrt{\frac{1}{2}}|8,-7\rangle\right) \otimes \label{destructive_interference_3}
        \\
    &\left(\frac{1}{2}|0,0\rangle-\frac{1}{2\sqrt{5}}|1,0\rangle
    -\frac{1}{2}|2,0\rangle+\frac{3}{2\sqrt{5}}|3,0\rangle\right) \label{destructive_interference_4}
        \\ 
    &+\alpha\gamma\left(\frac{2}{\sqrt{15}}|6,-6\rangle-\sqrt{\frac{1}{2}}|7,-6\rangle+\sqrt{\frac{7}{30}}|8,-6\rangle\right) \otimes
        \\
    &\left(\sqrt{\frac{3}{10}}|1,-1\rangle+\sqrt{\frac{1}{2}}|2,-1\rangle+\sqrt{\frac{1}{5}}|3,-1\rangle\right)
        \\ 
    &+\alpha\gamma\left(\frac{2}{\sqrt{15}}|6,-6\rangle+\sqrt{\frac{1}{2}}|7,-6\rangle+\sqrt{\frac{7}{30}}|8,-6\rangle\right) \otimes
        \\
    &\left(\sqrt{\frac{3}{10}}|1,-1\rangle-\sqrt{\frac{1}{2}}|2,-1\rangle+\sqrt{\frac{1}{5}}|3,-1\rangle\right)
        \\ 
    &+\beta\gamma\left(\frac{3}{\sqrt{26}}|5,-5\rangle-\sqrt{\frac{1}{10}}|6,-5\rangle-\sqrt{\frac{2}{13}}|7,-5\rangle+\sqrt{\frac{2}{5}}|8,-5\rangle\right) 
        \\
    &\otimes\left(\sqrt{\frac{1}{2}}|2,-2\rangle+\sqrt{\frac{1}{2}}|3,-2\rangle\right)
        \\ 
    &-\beta\gamma\left(\frac{3}{\sqrt{26}}|5,-5\rangle+\sqrt{\frac{1}{10}}|6,-5\rangle-\sqrt{\frac{2}{13}}|7,-5\rangle-\sqrt{\frac{2}{5}}|8,-5\rangle\right) 
        \\
    &\otimes\left(-\sqrt{\frac{1}{2}}|2,-2\rangle+\sqrt{\frac{1}{2}}|3,-2\rangle\right).
\end{align}
\endgroup

One can observe that in the $\alpha^{2}, \beta^{2}$, and $\gamma^{2}$ terms formed by KRb in the same state, only products with K$_2$ in symmetric nuclear spin states $\ket{S}$ and Rb$_2$ in symmetric nuclear spin states $\ket{S}$ are produced. On the other hand, the cross terms yield K$_2$ and Rb$_2$ product pairs in all combinations, including $\ket{A,A}$, $\ket{A,S}$, $\ket{S,A}$ and $\ket{S,S}$ nuclear spin states. When the reactants are prepared in the pure state, the coefficients of K$_2$ and Rb$_2$ in $\ket{A,S}$ and $\ket{S,A}$ nuclear spin states perfectly cancel from destructive interference. To demonstrate how this destructive interference happens, we can focus on Equation \ref{destructive_interference_1} to \ref{destructive_interference_4}. After summing up the terms, we find that the probability amplitudes of the $\ket{7,-7} \otimes \ket{1,0}$, $\ket{7,-7} \otimes \ket{3,0}$, $\ket{8,-7} \otimes \ket{0,0}$ and $\ket{8,-7} \otimes \ket{2,0}$ terms cancel, as their coefficients in Equation \ref{destructive_interference_1} to \ref{destructive_interference_2} and equation \ref{destructive_interference_3} to \ref{destructive_interference_4} have opposite signs. One can repeat this for other terms to show that only products in $\ket{S,S}$ and $\ket{A,A}$ states can be observed, with their probabilities given by \cite{hu2021nuclear}:

\begin{align}
&P_{AA} = |\alpha\beta|^{2} + |\beta\gamma|^{2} + |\alpha\gamma|^{2}, ~~~\text{and}
\\
&P_{SS} = 1 - P_{AA}.
\end{align}
At the magnetic field of B = 5 G, where $(\alpha,\beta,\gamma) = (0.595, 0.719, 0.359)$, we have $P_{AA} = 0.2951$ and $P_{SS} = 0.7049$. However, if the reactants are prepared in a fully mixed state $\rho_\mathrm{KRb} = |\alpha|^2 \ket{-4,1/2}\otimes\bra{-4,1/2} + |\beta|^2 \ket{-3,-1/2}\otimes\bra{-3,-1/2} + |\gamma|^2 \ket{-2,-3/2}\otimes\bra{-2,-3/2}$, populations add and interference between different KRb nuclear spin components vanishes. In this case, the probabilities to observe products in $\ket{A,A}$, $\ket{S,S}$, $\ket{A,S}$ and $\ket{S,A}$ states are:

\begin{align}
&P_{AA} =P_{AS}=P_{SA} = \frac{1}{2}\left(|\alpha\beta|^{2} + |\beta\gamma|^{2} + |\alpha\gamma|^{2}\right), ~~~\text{and}
\\
&P_{SS} = 1 - 3P_{AA}.
\end{align}

At the same magnetic field B = 5 G, the probability to observe product pairs in all four cases are: $P_{AA} = P_{AS} = P_{SA} = 0.1475$ and $P_{SS} = 0.5575$. We expect the formula for the full statistical mixture case to be valid whenever the coherence is completely lost before or during the reaction. Due to the spin-rotation correspondence, $[P_{eo},P_{ee},P_{oe},P_{oo}] = [P_{SS},P_{SA},P_{AA},P_{AS}]$ always holds, where $P_{eo}$ denotes the probability that $|N_{\mathrm{K}_2},N_{\mathrm{Rb}_2}\rangle=|even,odd\rangle$ and $P_{ee}$, $P_{oe}$ and $P_{oo}$ are defined similarly.

\subsection{Resonance-enhanced multiphoton ionization (REMPI) of reaction products}\label{appendix:rempi}

State-selective ionization of K$_{2}$ and Rb$_{2}$ reaction products is achieved through a resonance-enhanced multiphoton ionization scheme, which has been described in our previous works \cite{liu2021precision, hu2021nuclear}. We resonantly drive the K$_{2}$ and Rb$_{2}$ molecules through Q branch transitions from their initial states, $X^{1}\Sigma^{+}_{g}(\nu = 0, N_{\mathrm{K}_{2}})$ and $X^{1}\Sigma^{+}_{g}(\nu = 0, N_{\mathrm{Rb}_{2}})$, to intermediate electronic excited states, $B^{1}\Pi_{u}(\nu' = 1, N'_{\mathrm{K}_{2}}=N_{\mathrm{K}_{2}})$ and $B^{1}\Pi_{u}(\nu' = 4, N'_{\mathrm{Rb}_{2}}=N_{\mathrm{Rb}_{2}})$, using continuous-wave diode lasers at 648 nm and 674 nm that are stabilized to an accuracy of 10 MHz through locking to a calibrated wavelength meter. The 648~nm beam has a power of 14.5~mW and a 1/$e^{2}$ Gaussian beam waist of 750~$\mu$m at the molecule position. Similarly, the 674~nm beam has a power of 10.6~mW and a beam waist of 750~$\mu$m. Following excitation, the product molecules are ionized by a pulsed 532~nm laser with pulse energies of 200 $\mu$J. The resonant legs (648~nm and the 674~nm) are turned on simultaneously with a length of 100 ns, while the 532 nm pulse is triggered to arrive 20 ns after the on edge of the resonant legs. We note that our 532 nm laser pulse energy saturates the ionization transition, as no significant increase of ion signal can be observed with higher pulse energies. The ions are then accelerated under an electric field created by 6 field plates in a velocity map imaging (VMI) configuration~\cite{eppink1997velocity} and are collected by a microchannel plate (MCP) detector. 

For each experimental cycle, we start the REMPI ionization sequence after the creation and the initial state preparation of the reactant KRb molecules. We strobe the optical dipole trap (ODT) at 10 kHz with a 50\% duty cycle square wave pattern for 1~s, as shown in Fig.~\ref{fig:coherent}a, and fire the REMPI pulses at the end of every ODT dark period to probe the reaction products in the absence of the trapping light. We maintain the same time-averaged intensity for the ODT while strobed as while static. To ensure that the detected products do not include those formed when the ODT is on, a 2 $\mu\text{s}$ clean-up pulse consisting of the 648 and 674 nm beams is added at the beginning of each ODT dark period to remove the K$_2$ and Rb$_2$ products formed while the ODT is on. 

\subsection{Coincidence detection of reaction products}\label{appendix:coi}

\begin{figure*}
\centering
\includegraphics[width= \textwidth]{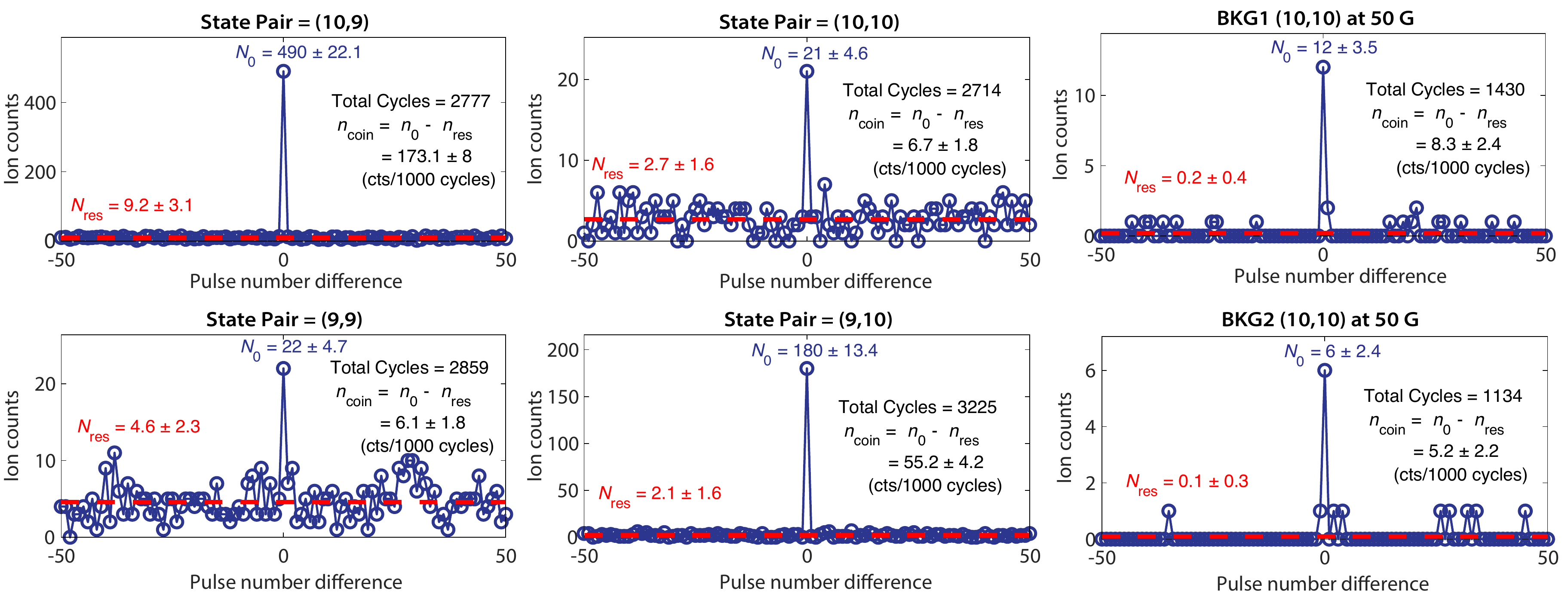}
\caption{\textbf{} Coherent reaction data at 5 G together with the background measurement at 50 G. K$_2$-Rb$_2$ ion pair numbers that pass the momentum and time-of-flight filtering as a function of the pulse number difference. Note that in Fig.~\ref{fig:coherent}b in the main figure, the y-axes are normalized to per 1000 cycles.}
\label{fig:raw_data}
\end{figure*}

Coincidence detection of the K$_{2}$ and Rb$_{2}$ product molecules created in the same reaction event is necessary both to study correlations of products and to allow comparison of our data with models that distinguish whether coherence is preserved throughout the chemical reactions or not. Our implementation involves the simultaneous REMPI of K$_2$ and Rb$_2$ followed by ion data filtering using momentum conservation. The analysis takes three steps. We first apply a coarse filtering by selecting the K$_2$ and Rb$_2$ pairs coming from the same ionization pulse as shown in Fig.~\ref{fig:coherent}a. Next, we apply momentum conservation criteria $P_{i, \mathrm{K}_{2}} + P_{i, \mathrm{Rb}_{2}} = 0$, where $i = (x, y, z)$ on the selected K$_2$ and Rb$_2$ pairs. Finally, we subtract the false positive residual coincidence signals ($N_{res}$) after the second step. Fig.~\ref{fig:raw_data} shows the data for the four state pairs investigated at 5 G as well as the background measurements of the (10,10) state pair at 50 G (repeated twice). The details of the analysis are the following. 

The momentum filtering is achieved using two observables: the impact positions on the MCP detector and the arrival time-of-flights (TOF). In our experiment setup, we operate the electrode plates responsible for accelerating the product ions in a VMI configuration, which maps the transverse momenta ($P_{x,\mathrm{K}_{2}}, P_{y,\mathrm{K}_{2}}$) and ($P_{x,\mathrm{Rb}_{2}}, P_{y,\mathrm{Rb}_{2}}$) onto the impact positions ($X_{\mathrm{K}_{2}}, Y_{\mathrm{K}_{2}}$) and ($X_{\mathrm{Rb}_{2}}, Y_{\mathrm{Rb}_{2}}$) on the MCP detector. For each ion species $s$ ($s = \mathrm{K}_{2}$ or $\mathrm{Rb}_{2}$), the positions on MCP detectors are $X_{s} = X_{0,s} + A\sqrt{2M_{s}}P_{x,s}$ and $Y_{s} = Y_{0,s} + A\sqrt{2M_{s}}P_{y,s}$. Here, $X_{0,s}$, $Y_{0,s}$ and $A$ are determined by the magnetic field, the acceleration voltage and the geometry of the electrode plates during ionization, which are kept constant between experimental cycles. Meanwhile, the momenta along the TOF axis ($P_{z,\mathrm{K}_{2}}, P_{z,\mathrm{Rb}_{2}}$) are mapped onto the total TOFs, that are given by $P_{z,s} = (TOF_{s} - TOF_{0,s})/(\eta \delta t/TOF_{0,s} - 1)$. Here, $TOF_{0,s}$ represents the center of the $TOF_s$ distribution and is determined by the acceleration voltages. The parameter $\delta t$ represents the free-flying time of neutral products after the chemical reaction but before ionization, with a uniform distribution ranging from 0 to 45 $\mu$s. In our system, $\eta$ depends on the electrode geometry and is equal to 136 \cite{Liu2020}. This formula accounts for the ionization of product molecules occurring at different positions between the repeller and extractor plates.

We set our coincidence detection filter criteria that satisfies momentum conservation to be:

\begin{align}
    &|(X_{\mathrm{K}_{2}} - X_{0,\mathrm{K}_{2}}) + \sqrt{M_{\mathrm{Rb}_{2}}/M_{\mathrm{K}_{2}}}(X_{\mathrm{Rb}_{2}} - X_{0,\mathrm{Rb}_{2}})| \leq n\sigma_{X}\sqrt{1 + M_{\mathrm{Rb}_{2}}/M_{\mathrm{K}_{2}}},
    \\
    &|(Y_{\mathrm{K}_{2}} - Y_{0,\mathrm{K}_{2}}) + \sqrt{M_{\mathrm{Rb}_{2}}/M_{\mathrm{K}_{2}}}(Y_{\mathrm{Rb}_{2}} - Y_{0,\mathrm{Rb}_{2}})| \leq n\sigma_{Y} \sqrt{1 + M_{\mathrm{Rb}_{2}}/M_{\mathrm{K}_{2}}},~~~~\text{and}
    \\
    &|(TOF_{\mathrm{K}_{2}} - TOF_{0,\mathrm{K}_{2}}) + (TOF_{\mathrm{Rb}_{2}} - TOF_{0,\mathrm{Rb}_{2}}) \frac{(\eta \delta t/TOF_{0,\mathrm{K}_{2}} - 1)}{(\eta \delta t/TOF_{0,\mathrm{Rb}_{2}} - 1)}| \leq n_{T}\sigma_{T} \sqrt{1 + (\frac{(\eta \delta t/TOF_{0,\mathrm{K}_{2}} - 1)}{(\eta \delta t/TOF_{0,\mathrm{Rb}_{2}} - 1)})^{2}}.
\end{align}
Here, $\sigma_{X}$, $\sigma_{Y}$, and $\sigma_{T}$ are the 1$\sigma$ resolution of the detection system, which depends on the acceleration voltage. When $V_{rep} = 100$ V, which is used in the 5 G coherent reaction data (Fig.~\ref{fig:coherent} in main text), we have $\sigma_{X} = 0.23$ mm, $\sigma_{Y} = 0.23$ mm, and $\sigma_{T} = 11$ ns. We use $V_{rep} = 2000$ V for the statistical mixture data as shown in Fig.~\ref{fig:incoherent} in the main text. When $V_{rep} = 2000$ V, $\sigma_{X} = 0.08$ mm, $\sigma_{Y} = 0.08$ mm, and $\sigma_{T} = 4$ ns. The choice of $n=3$ and $n_{T}=3$ are determined empirically such that no significant loss of coincidence signal occurs while the noise background is sufficiently suppressed. We note that the choice of the multiplicative factors $n$ and $n_{T}$ does not affect the population ratios between different product rotation pairs significantly.

\begin{figure*}
\centering
\includegraphics[width= \textwidth]{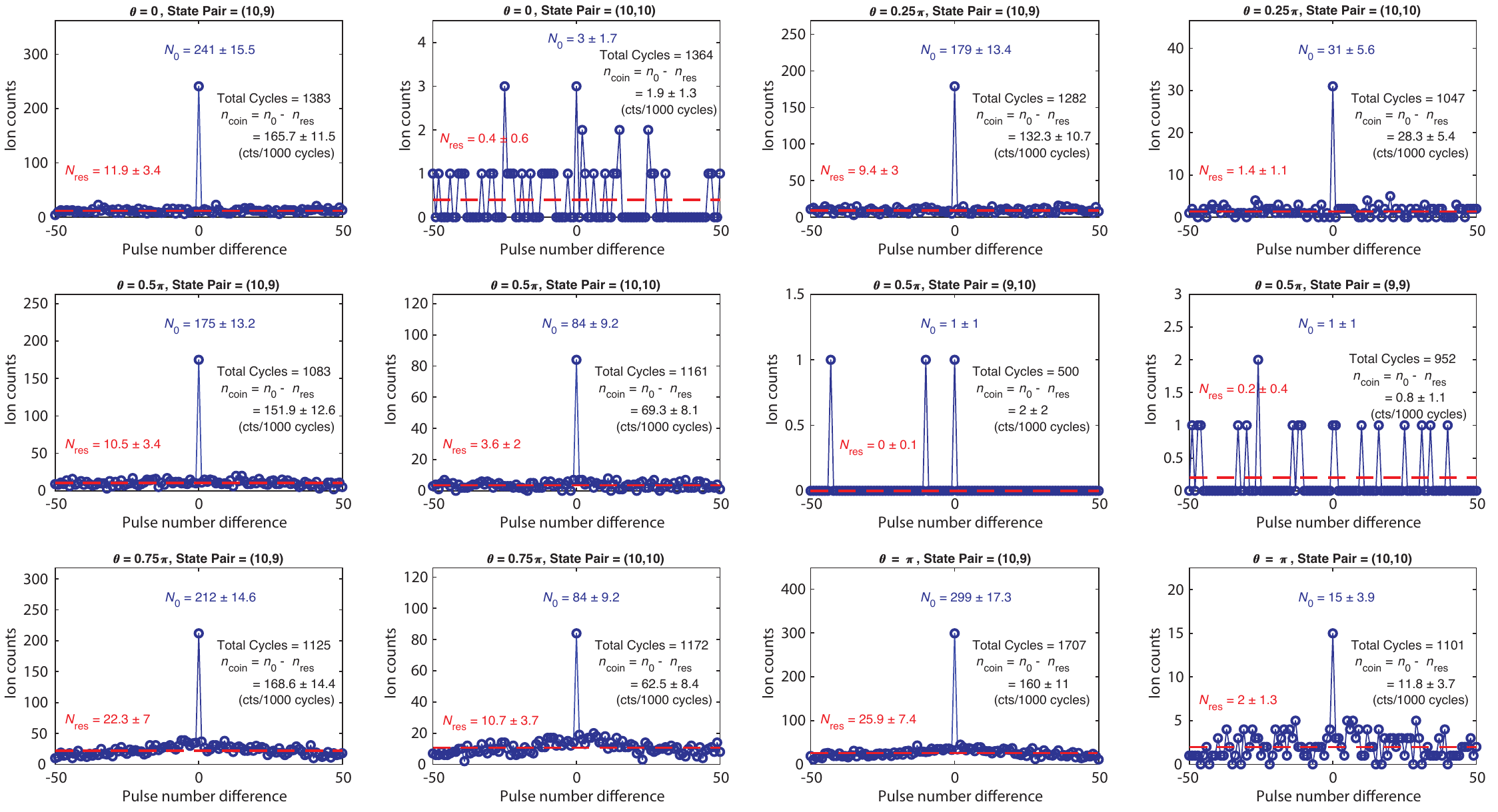}
\caption{\textbf{} Statistical mixture data with different up-leg rotation angle $\theta$. K$_2$-Rb$_2$ ion pair numbers that pass the momentum and time-of-flight filtering as a function of the pulse number difference for different state pairs.}
\label{fig:raw_data_mixture}
\end{figure*}

Even with the transverse momentum and TOF filters, there are still some false positive coincidence signals (non-coincidence counts) from two independent reaction events detected within a single REMPI pulse that happen to satisfy the transverse momentum and TOF filtering criteria. In order to identify the level of these non-coincidence counts, we offset the relative K$_2$ and Rb$_2$ pulse number and apply the same transverse momentum and TOF filters. We change the pulse number difference in the range of [-50, 50]. In Fig.~\ref{fig:raw_data}, we plot the number of K$_2$-Rb$_2$ pairs after the momentum filtering as a function of pulse number difference for the different state pairs. We average the ion counts with non-zero pulse number difference to get $N_{res}$, as indicated by the red dashed lines in Fig.~\ref{fig:raw_data}. $N_0$ is the number of K$_2$-Rb$_2$ ion pairs from the same pulse that pass the TOF and the X, Y momentum filters. In the main text, we report the coincidence counts normalized to per 1000 cycles $n_{coin} = (N_0-N_{res})\times 1000/N_{c}$, where $N_{c}$ is the total number of cycles. The statistical mixture data is shown in Fig.~\ref{fig:raw_data_mixture}, which corresponds to the data in Fig.~\ref{fig:incoherent} in the main text.

\subsection{Preparation of entangled KRb state by magnetic field ramping}\label{appendix:Hamiltonian}

In the presence of an external magnetic field, the molecular Hamiltonian in the electronic, vibrational ground state is \cite{aldegunde2008hyperfine, brown2003rotational}:

\begin{align}
    H_{\mathrm{KRb}} = H_{\mathrm{rot}} + H_{\mathrm{hyperfine}} + H_{\mathrm{Zeeman}}
\end{align}

\begin{figure*}
\centering
\includegraphics[width= \textwidth]{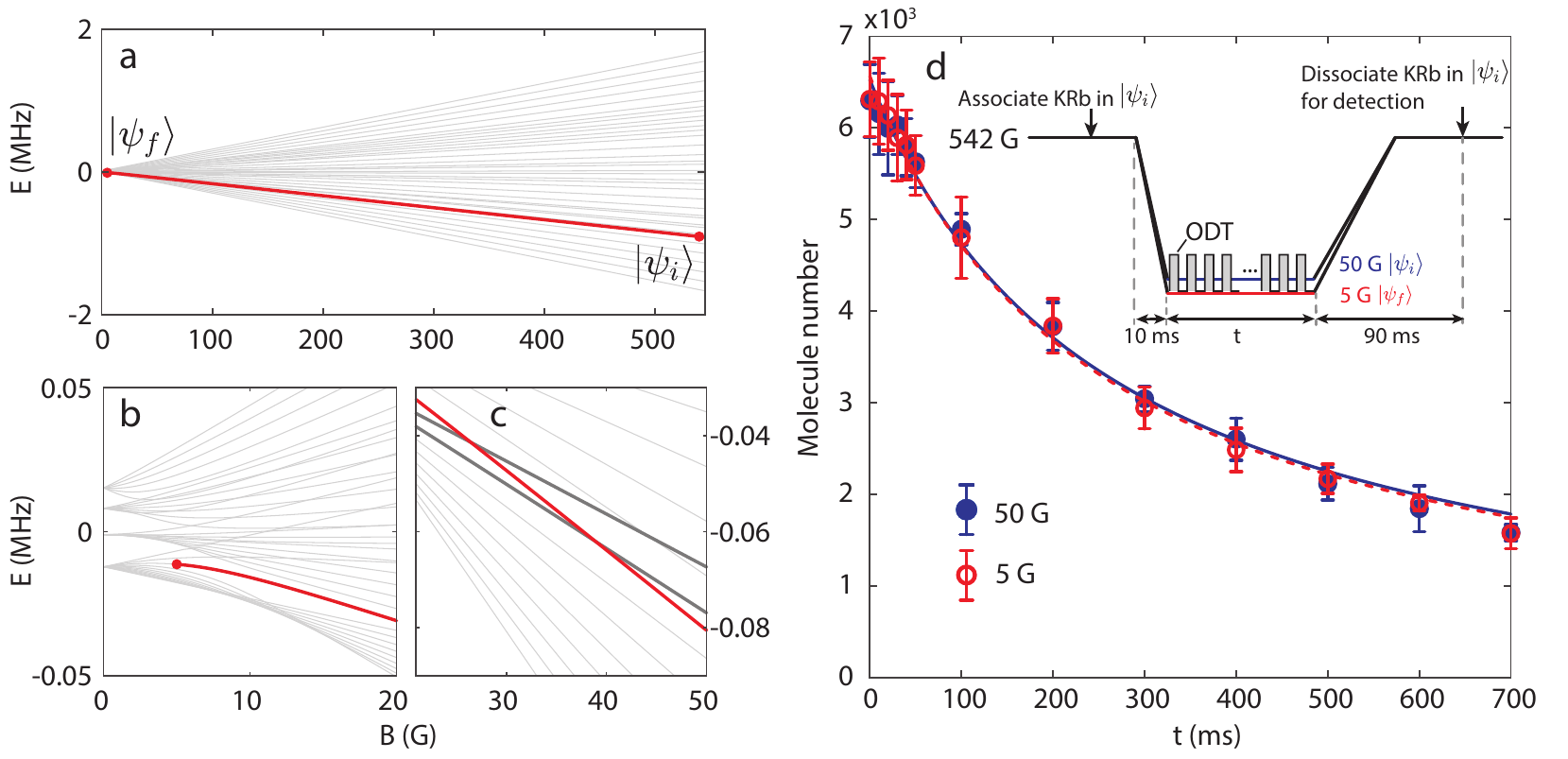}
\caption{\textbf{} \textbf{a-c,} KRb molecule eigenenergies as a function of magnetic field. \textbf{d,} Molecule lifetime characterization under ODT modulation and electric field 17 V/cm at 5 G (open red) and 50 G (solid blue). Error bars represent the standard deviation of 10 measurements. The solid (dashed) line is a fit to a two-body decay term at 50 G (5 G) respectively. The fitted half-lives are 237(20) ms and 243(12) ms for 50 G and 5 G respectively. The fitting error represents a 95\% confidence interval. The inset shows the magnetic field ramping timing scheme.}
\label{fig:molecule_lifetime}
\end{figure*}

Where,

\begin{align}
    &H_{\mathrm{rot}} = B_{\mathrm{rot}} \textbf{N}^{2},
    \\
    &H_{\mathrm{hyperfine}} = c_{\mathrm{K}} \textbf{N} \cdot \textbf{I}_{\mathrm{K}} + c_{\mathrm{Rb}} \textbf{N} \cdot \textbf{I}_{\mathrm{Rb}} + c_{4} \textbf{I}_{\mathrm{K}} \cdot \textbf{I}_{\mathrm{Rb}} + \textbf{V}_{\mathrm{K}} \cdot \textbf{Q}_{\mathrm{K}} + \textbf{V}_{\mathrm{Rb}} \cdot \textbf{Q}_{\mathrm{Rb}},~~~\text{and}
    \\
    &H_{\mathrm{Zeeman}} = -g_{\mathrm{R}}\mu_{\mathrm{N}} \textbf{N} \cdot \textbf{B} - g_{\mathrm{K}}\mu_{\mathrm{N}} (1-\sigma_{\mathrm{K}}) \textbf{I}_{\mathrm{K}} \cdot \textbf{B} - g_{\mathrm{Rb}}\mu_{\mathrm{N}} (1-\sigma_{\mathrm{Rb}}) \textbf{I}_{\mathrm{Rb}} \cdot \textbf{B}.
\end{align}
Here, $B_{\mathrm{rot}}$ and $\textbf{N}$ are the rotational constant and rotational angular momentum operator for the product $\mathrm{K}_{2}$ or $\mathrm{Rb}_{2}$ molecules, $\textbf{I}_{\mathrm{K}}$ and $\textbf{I}_{\mathrm{Rb}}$ are the nuclear spin angular momentum operators of K and Rb in the molecule, $\textbf{V}_{\mathrm{K}}$ and $\textbf{V}_{\mathrm{Rb}}$ are the intramolecular electric field gradients at the K and Rb nuclei, $\textbf{Q}_{\mathrm{K}}$ and $\textbf{Q}_{\mathrm{Rb}}$ are the nuclear electric quadruple tensor operators of the K and Rb nuclei, $g_{\mathrm{R}}$, $g_{\mathrm{K}}$ and $g_{\mathrm{Rb}}$ are the rotational g-factors of the molecule and the individual nuclear g-factors of K and Rb, $\mu_{\mathrm{N}}$ is the nuclear magneton, $\sigma_{\mathrm{K}}$ and $\sigma_{\mathrm{Rb}}$ are the nuclear shielding constants of K and Rb, and \textbf{B} represents the external magnetic field. Using literature values \cite{ni2008high,aldegunde2008hyperfine, neyenhuis2012anisotropic} for KRb molecules, we can calculate the energy eigenstates in the $\ket{m_{\mathrm{I}}^{\mathrm{K}},m_{\mathrm{I}}^{\mathrm{Rb}}}$ basis at arbitrary magnetic fields, which are shown in Fig.~\ref{fig:molecule_lifetime}a as well as their state decomposition.

In our experiment, we create an entangled nuclear spin state using a magnetic field ramp from B = 542 G to B = 5 G within 10 ms. The initial KRb state at the start of this ramp is $\ket{\psi_{i}} = \ket{0,0,-4,1/2}$ at B = 542 G. Throughout the field ramp, the KRb molecules remain in an energy eigenstate following the red curve shown in Fig.~\ref{fig:molecule_lifetime}a-c and end in $\ket{\psi_{f}} = 0.595\ket{0,0,-4,1/2} + 0.719\ket{0,0,-3,-1/2} + 0.359\ket{0,0,-2,-3/2}$ at B = 5 G. We note that the field ramp is sufficiently fast through avoided crossings from states with different total nuclear spin projections, so the molecules stay in the eigenstate with the same total nuclear projection, which connects to $|I_\mathrm{tot}=\frac{11}{2},M_{I_\mathrm{tot}} = {-\frac{7}{2}}\rangle$ at zero magnetic field. To check that the KRb reactants stay coherent before the reaction, we measure the lifetime of state $\ket{\psi_f}$ at 5 G under ODT modulation and a constant electric field 17 V/cm. We ramp the magnetic field back up to 542 G and dissociate $|0,0,-4,1/2\rangle$ KRb molecules into K and Rb atoms for detection [Fig.~\ref{fig:molecule_lifetime}d (inset)]. If there is phase randomization between different spin components or if the state's decomposition changes, the KRb will not end up in $\ket{\psi_i}$ after ramping the magnetic field up to 542 G, resulting in apparent molecule loss, as the molecule detection is sensitive to the molecule hyperfine state.
Also, if decoherence happens and molecules become distinguishable, the molecules no longer experience a p-wave barrier, allowing for reactions in the much faster s-wave channel. We measure the molecule lifetime at 50 G, where the KRb state is in $\ket{0,0,-4,1/2}$ (with 0.994 population), as a reference to account for losses due to the field ramp. There is no observable difference between the two lifetimes as shown in Fig.~\ref{fig:molecule_lifetime}d, implying that the molecules stay coherent before reacting. 

\subsection{Normalization of coincident counts}\label{appendix:normalization}

To correctly extract the populations from the number of coincidence counts measured in the $N_{\mathrm{K}_{2}}$ = 9, 10 and $N_{\mathrm{Rb}_{2}}$ = 9, 10 channels, we normalize our data to account for detection biases. The two major biases are the velocity-dependent detection efficiency $C$ and the product state degeneracy $D$. The number of expected coincidence ion counts for state $\ket{i,j}$ is proportional to $P_{ij}C_{ij}D_{ij}$, where $P_{ij}$ is the branching ratio we aim to extract and compare with the theoretical prediction.

The dependence of the detection efficiency on the product velocity comes from: (1) a product with higher velocity is more likely to escape the detection region covered by the REMPI beams before they are pulsed on and (2) a product with higher velocity along the propagation direction of the REMPI beams experiences a larger Doppler shift of the bound-to-bound transition, which affects the probability of promoting the molecule into the excited state. We perform a Monte Carlo simulation to get the normalization factors accounting for these two velocity dependencies. 

In the simulation, we sample $10^5$ K$_2$-Rb$_2$ pairs in each state and calculate the total coincidence counts of each state. The magnitudes of the velocities for all relevant state pairs are calculated using energy conservation, the previously reported total exoergicity of 9.77 cm$^{-1}$, and the rotational constants of $^{40}\mathrm{K}_{2}$ and $^{87}\mathrm{Rb}_{2}$~\cite{liu2021precision}. We calculate the total kinetic energy of product pairs for all four relevant rotation states as: E($N_{\mathrm{K}_{2}}$ = 9, $N_{\mathrm{Rb}_{2}}$ = 9) = 2.871 cm$^{-1}$, E($N_{\mathrm{K}_{2}}$ = 10, $N_{\mathrm{Rb}_{2}}$ = 9) = 1.776 cm$^{-1}$, E($N_{\mathrm{K}_{2}}$ = 9, $N_{\mathrm{Rb}_{2}}$ = 10) = 2.434 cm$^{-1}$ and E($N_{\mathrm{K}_{2}}$ = 10, $N_{\mathrm{Rb}_{2}}$ = 10) = 1.338 cm$^{-1}$. The corresponding product pair velocities are listed in Table.~\ref{table:calibration}. 
For each K$_2$-Rb$_2$ pair, the K$_2$ velocity direction is sampled from a uniform spherical distribution and the Rb$_2$ flies towards the opposite direction as K$_2$. 

The creation time of each product pair is uniformly sampled from t = 0 to t = 45 $\mathrm{\mu}$s to represent the random starting time of each reaction event. The starting locations of all products are the same and are set to the center of the REMPI beams. Then we track the position of each product molecule at t = 45 $\mathrm{\mu}$s when the REMPI beams turn on. 

We model the resonant excitation in the REMPI process using a two-level system with decay in the excited state~\cite{liu2021precision}. The dynamics of the ground state $|0\rangle$ in the X$^1\Sigma$ manifold and the intermediate state $|1\rangle$ in the B$^1\Pi$ can be described using the following rate equations
\begin{align}
\frac{d}{dt} \rho_{00} & = -\frac{i}{2}(\Omega_{01}\rho_{10}-c.c.), \\
\frac{d}{dt} \rho_{11} & = -\Gamma\rho_{11} + \frac{i}{2} (\Omega_{01}\rho_{10}-c.c.), \\
\frac{d}{dt} \rho_{10} & = -\frac{1}{2}\Gamma \rho_{10} + i \Delta \rho_{10} + \frac{i}{2} \Omega_{10}(\rho_{11}-\rho_{00}), 
\end{align}
where $c.c.$ stands for complex conjugate, and $\Omega_{01}$ is the Rabi rate of the bound-to-bound transition. $\Delta$ is the 648 (674) nm laser detuning, which comes from the Doppler shift for each molecule. $\Gamma$ is the decay rate of the excited state $|1\rangle$ and we use (2$\pi$)14 MHz for both K$_2$ and Rb$_2$~\cite{liu2021precision}. Here, $\Omega_{01}$ is position dependent with a peak value of (2$\pi$)15 MHz and decays following a Gaussian profile with a 1/$e^{2}$ beam waist of 750 $\mathrm{\mu}$m. Note that we use the same Rabi rate for all different initial rotational states, as the Q branch rotational transition dipole moments are the same for different initial rotational states~\cite{hu2021nuclear,hansson2005comment}. In the experiment, we cast a disk shape dark mask on the molecule cloud to protect the reactants from the REMPI beams. The inner dark region has a radius of 125 $\mu$m and the beam is cut off at a 1 mm radius. We incorporate these geometric factors in the simulation as well. 

We record the excited state population $\rho_{11,\mathrm{K_2}}$ and $\rho_{11,\mathrm{Rb_2}}$ 20 ns after the 648 nm or the 674 nm beam is turned on when the 532 nm laser pulse is fired. We assume that the 532 ionization rate is constant and the final K$_2$(Rb$_2$) ion count is proportional to $\rho_{11,\mathrm{K_2}}$($\rho_{11,\mathrm{Rb_2}}$). Finally, we sum up the excited state population product $\rho_{11,\mathrm{K_2}}\cdot\rho_{11,\mathrm{Rb_2}}$ of each ionization event and get the coincidence counts of each state pair. The velocity-dependent calibration factor $C_{ij}$ is the coincidence counts of each state pair after normalizing to the total coincidence counts of the four states. We repeat the simulations ten times to obtain the averaged factor and the standard deviation. Note that we ignore the spatial dependence of the 532 nm beam ionization efficiency of the excited products, which only changes the result by less than 1\%. We also vary the Rabi rates of each resonant transition in a $\pm$(2$\pi$)5 MHz range which changed the calibration factor by $\sim$ 2\%. 

In addition, we assume that reaction dynamics in the rotational degree of freedom are completely statistical. The degeneracy of each product pair follows a simple counting model~\cite{liu2021precision}. Given $N_\mathrm{K_2}$ and $N_\mathrm{Rb_2}$, the degeneracy is calculated by counting the number of different ways of coupling three angular momenta of the products, namely the rotation of K$_2$ ($\mathbf{N}_{\mathrm{K_2}}$), the rotation of Rb$_2$ ($\mathbf{N}_{\mathrm{Rb_2}}$), and the relative rotation between K$_2$ and Rb$_2$ ($\mathbf{L}$), that result in a total angular momentum $\mathbf{J}_{\mathrm{tot}} = \mathbf{N}_{\mathrm{K_2}}+\mathbf{N}_{\mathrm{Rb_2}} + \mathbf{L} = 1$. We assume that $\mathbf{J}_{\mathrm{tot}}$ is conserved. Initially, $\mathbf{J}_{\mathrm{tot}} = 1$ because the molecules are indistinguishable fermions and hence react through p-wave collisions. The resulting $D_{ij}$ and the final total detection coefficient $c_{ij} = C_{ij}D_{ij}$ are listed in Table~\ref{table:calibration}.

\begin{table}
\captionsetup{justification=centering}
\begin{center}
\begin{tabular}{c | r r r r  }
Description & $|10,9\rangle$  & $|10,10\rangle$  & $|9,10\rangle$ & $|9,9\rangle$  \\
\hline
velocities $(v_\mathrm{K_2},v_\mathrm{Rb_2})$ (m/s) & (19.1,8.8)  & (16.6,7.6) & (22.3,10.3) & (24.3,11.2) \\
velocity-dependent calibration factor $C$ &  0.2757(9)  & 0.3447(6) & 0.2055(6) &0.1741(8)  \\
channel degeneracy $D$ & 29 &31 &29 &28\\
total detection coefficient $c$ & 0.2709(8)  & 0.3621(6) & 0.2019(6) & 0.1651(8)
\end{tabular}
\end{center}
\caption{Velocity dependent calibration factors for each state pair.}
\label{table:calibration}
\end{table}

\subsection{Bayesian analysis of coherence ($\Gamma$)}\label{appendix:bayesian}

Estimation of coherence ($\Gamma$) from the measurements is complicated by the fact that $\Gamma$ is near its maximum value of one, causing a description involving Gaussian distributions to be cut off and unsuitable. Instead, we apply Bayesian analysis to derive the probability distribution for $\Gamma$ given the six measured counts $N_{0,j}$ ($j=1...6$) and their associated adjustment factors. 

Each measurement samples a Poisson distribution 
\begin{equation}
f(N_{0,j}|\lambda_j(\Gamma,\mu))=\frac{e^{-\lambda_j} \lambda_j ^{N_{0,j}}}{N_{0,j}!},
\end{equation} where the expected counts $\lambda_j$ depend on the coherence parameter $\Gamma$, the background rate $\mu$, and various known values describing the experiment. The probability of measuring our data given parameters $\Gamma,\mu$ is then
\begin{equation}
p(N_{0}|\Gamma,\mu)=\prod^6_{j=1}f(N_{0,j}|\lambda_j(\Gamma,\mu)),
\end{equation}
with expected counts 
\begin{equation}
\lambda_j(\Gamma,\mu) = 10^{-7} M_j N_{c,j} \left(\alpha c_j \left( P_{C,j} \Gamma + (1-\Gamma)P_{S,j}\right)+\mu\right)+N_{res,j}.
\end{equation}
Symbol values are listed in Table \ref{table:bayes} and $\alpha$ is an event rate such that
\begin{equation}
\alpha = \sum _{j=1}^4 \left(\frac{N_{0,j}-N_{res,j}}{10^{-7} M_j N_{c,j}}-\mu\right)/c_j 
 = \sum _{j=1}^4 \left(\frac{\lambda_j-N_{res,j}}{10^{-7} M_j N_{c,j}}-\mu\right)/c_j
\end{equation}
as the total of all branching ratios is 1:
\begin{equation}
 \sum _{j=1}^4 P_{C,j} = \sum _{j=1}^4 P_{S,j} = 1.
\end{equation}

\begin{figure}
\centering
\includegraphics[width=0.5\textwidth]{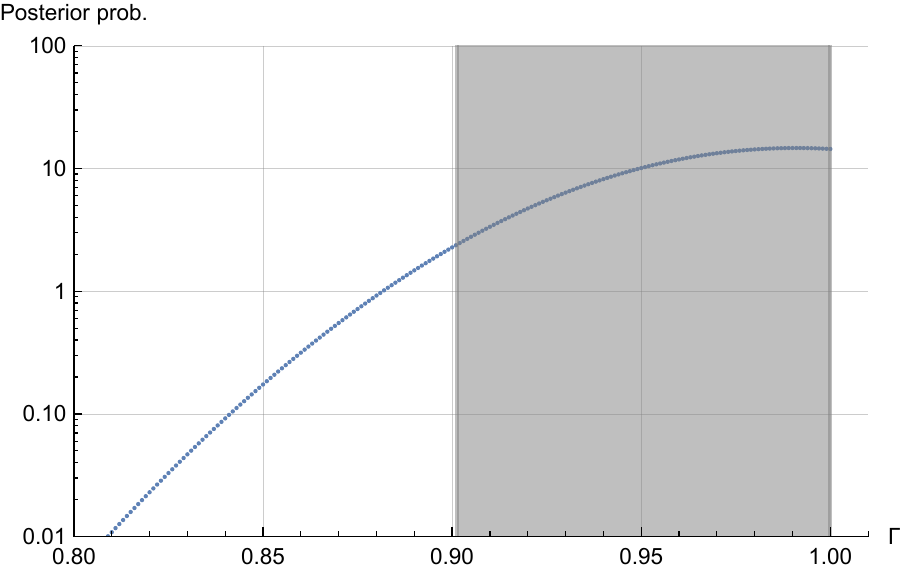}
\caption{Probability density function $p(\Gamma)$. The 95\% confidence interval $L$=0.9014 to 1 is shaded, derived from the limit of integration for which $\int^1_L p(\Gamma)d\Gamma=0.95$.}
\label{figure:Gamma} 
\end{figure}

By application of Bayes' theorem~\cite{Wasserman} under the assumption of a flat prior where all pairs $(\Gamma,\mu)$ are equally likely:
\begin{equation}
p(\Gamma,\mu)=\frac{p(N_{0}|\Gamma,\mu)}{\int p(N_{0}|\Gamma,\mu) d\Gamma d\mu}
\end{equation}
This joint probability density is integrated over $\mu$ to find the probability distribution for $\Gamma$, as shown if Fig.~\ref{figure:Gamma}:
\begin{equation}
p(\Gamma)=\int_0^\infty p(\Gamma,\mu) d\mu
\end{equation}

\begin{table}
\begin{tabular}{c | c | r r r r r r }
Description & Symbol & $\ket{10,9}$  & $\ket{10,10}$ & $\ket{9,10}$ & $\ket{9,9}$ & BKG1 & BKG2 \\
\hline
index & $j$ & 1 & 2 & 3 & 4 & 5 & 6 \\
observed counts & $N_0$ & 490  & 21 & 180 & 22 & 12 & 6 \\
expected counts & $\hat{\lambda}_j$ & 477.5  & 20.5 & 188.9 & 23.8 & 9.2 & 6.7 \\
molecule \# & $M$ & 11519  & 11143  & 11733 & 11801 & 11420 & 10484 \\
cycles & $N_c$ & 2777  & 2714 & 3225 & 2859 & 1430 & 1134 \\
residual baseline & $N_{res}$ & 9.2  & 2.7 & 2.1 & 4.6 & 0.2 & 0.1 \\
detection coeff. & $c$ & 0.2709  & 0.3621 & 0.2019 & 0.1651 & 0 & 0 \\
coherent branching & $P_C$ & 0.7049  & 0 & 0.2951 & 0 & 0 & 0 \\
statistical branching & $P_S$ & 0.5575 & 0.1475 & 0.1475 & 0.1475 & 0 & 0
\end{tabular}
\caption{Data and parameters for the Bayesian calculation of the probability distribution for $\Gamma$. For reference, expected counts $\hat{\lambda}_j$ are listed at the point of maximum likelihood ($\hat{\Gamma}=0.9957, \hat{\mu}=5.54; \hat{\alpha}=738.92$). Jeffreys' non-informative prior $\mu^{-1/2}$ was also tested, with negligible effect compared to the flat prior.}
\label{table:bayes}
\end{table}

\newpage

\end{document}